\documentclass[conference,romanappendices,10pt]{IEEEtran}	

\IEEEoverridecommandlockouts

\usepackage{algorithm,algorithmic,amsmath,amssymb,amsthm,array,bibentry,caption,cite,color,comment,enumerate}
\usepackage{enumitem,eurosym,float,graphicx,lettrine,mathrsfs,multicol,multirow,nomencl,pict2e,psfrag}
\usepackage{ragged2e,setspace,subfigure,supertabular,tabularx,url}
\usepackage[USenglish]{babel}

{		\newtheorem{theorem}{Theorem}}
{ 		}
{ 		\newtheorem{lemma}{Lemma}}
{ 		}
{ 		}
{		}
{		}
{ 		}

\newcommand{\p}{\mathbf{p}}

\newcommand{\setD}{\mathcal{D}}

\newcommand{\setH}{\mathcal{H}}
\newcommand{\setI}{\mathcal{I}}

\newcommand{\setU}{\mathcal{U}}

\newcommand{\argmax}{\operatornamewithlimits{argmax}}

\newcommand{\diff}{\mathrm{d}}
\newcommand{\Exp}{\mathbb{E}}

\renewcommand{\Pr}{\mathbb{P}}

\begin{document}
\title{User Scheduling and Optimal Power Allocation\\ for Full-Duplex Cellular Networks \vspace{-1mm}}
		
\author{
\IEEEauthorblockN{George~C.~Alexandropoulos, Marios Kountouris, and Italo Atzeni}
\IEEEauthorblockA{Mathematical and Algorithmic Sciences Lab, France Research Center, Huawei Technologies Co$.$ Ltd$.$}
emails: \{george.alexandropoulos, marios.kountouris, italo.atzeni\}@huawei.com \vspace{-3mm}}

\maketitle

\begin{abstract}
The problem of user scheduling and power allocation in full-duplex (FD) cellular networks is considered, where a FD base station communicates simultaneously with one half-duplex (HD) user on each downlink and uplink channel. First, we propose low complexity user scheduling algorithms aiming at maximizing the sum rate of the considered FD system. Second, we derive the optimal power allocation for the two communication links, which is then exploited to introduce efficient metrics for FD/HD mode switching in the scheduling procedure, in order to further boost the system rate performance. We analyze the average sum rate of the proposed algorithms over Rayleigh fading and provide closed-form expressions. Our representative performance evaluation results for the algorithms with and without optimal power control offer useful insights on the interplay among rate, transmit powers, self-interference (SI) cancellation capability, and available number of users in the system.
\end{abstract}


\section{Introduction} \label{sec:Intro}
\noindent Full-duplex (FD) communication is an emerging technology that has been recognized as one of the promising solutions to cope with the ever-growing demand for high data rates. FD systems allow simultaneous transmission and reception at the same time/frequency resource and have the potential to increase (theoretically double) the throughput in future (5G) wireless networks. There are two major challenges hindering the implementation of FD systems. First, the uplink (UL) communication is affected by the self-interference (SI) at the base station (BS), which is due to the signal leakage and imperfect isolation between transmit and receive antennas \cite{Dua12}. Second, the downlink (DL) communication suffers from the inter-user interference resulting from the UL mobile terminal (MT) using the same time/frequency resource.

Although resource allocation in half-duplex (HD) systems has been extensively studied, the new characteristics and challenges with FD communication have not been investigated yet. Without carefully allocating resources and selecting users, FD communication may cause excessive interference in both UL and DL, which may greatly limit the potential FD gains \cite{Goy15}. In this regard, \cite{Wan15} presents a distributed power allocation scheme for multi-cell FD scenarios that aims at maximizing the network throughput. A dynamic power control scheme in FD bidirectional networks was provided in \cite{Che13}. In \cite{Sul15}, a joint resource allocation problem including user pairing, subcarrier allocation, and power control for FD orthogonal division multiple access networks was considered. Low complexity user pairing algorithms that maximize the cell throughput or minimize the outage probability were presented in \cite{Cho14}. 
\begin{figure}[t!] 
\centering
\includegraphics[scale=0.8]{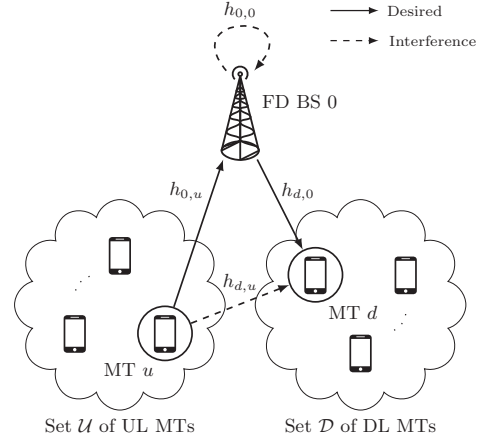} 
\caption{A small-cell FD cellular system with the desired and interfering channels at the receiving nodes.}
\label{Fig:scheme} \vspace{-3mm}
\end{figure}

In this work, we consider a FD BS communicating with one HD MT in the DL and one HD MT in the UL. We propose three low complexity user scheduling algorithms in order to maximize the system rate performance. In order to avoid the high complexity of inherent joint UL and DL scheduling, the selection procedure is decoupled and performed in two steps using efficient metrics. Furthermore, observing that for certain channel realizations, the HD mode in time-division duplex (TDD) outperforms the FD mode in terms of throughput, we derive a FD/HD switching method to further enhance the system rate. The switching metric is inspired by the optimal power allocation, which turns out to have a binary feature (e$.$g$.$, as in \cite{Gje08} for the $2$-user interference channel). In addition, we provide closed-form expressions for the average sum rate of the proposed algorithms over Rayleigh fading channels. The main takeaway of this paper is that power control and low complexity user scheduling can boost the performance of FD small/pico cells \cite{Atz15a} for realistic SI cancellation capabilities. 

\section{System Model} \label{sec:SM}
Consider the scenario illustrated in Fig.~\ref{Fig:scheme}, in which a small-cell FD BS communicates with a set $\setU \triangleq \{1,2,\ldots,K_\setU\}$ of HD MTs in the UL and a set $\setD \triangleq \{1,2,\ldots,K_\setD\}$ of DL HD MTs. In our system setup, the FD BS is indexed with $0$ and is equipped with one transmit and one receive antenna, whereas each MT has a single antenna. During a given time/ frequency resource unit, the BS schedules concurrently one MT from $\setU$ and one MT from $\setD$ so as to maximize the sum-rate performance; the selected UL and DL users are denoted as $u^{\star}\in\setU$ and $d^{\star}\in\setD$, respectively. The transmit powers of the BS and of every UL MT are denoted as $p_{0} \in [0, P_{0}]$ and $p_{\setU} \in [0, P_{\setU}]$, respectively, where $P_{0}$ and $P_{\setU}$ represent the maximum powers for the DL and UL communications, respectively. 

The narrowband complex-valued channel coefficient between a transmitting node $\ell\in\setU\cup\{0\}$ and a receiving node $k\in\setD\cup\{0\}$ is represented by $h_{k,\ell}$ and may include - in the general case - different propagation phenomena such as pathloss attenuation, small-scale fading, and shadowing. More specifically, $h_{0,u}$ represents the UL channel (i$.$e$.$, between the BS $0$ and the $u$-th UL MT), $h_{d,0}$ denotes the DL channel (i$.$e$.$, between the $d$-th DL MT and the BS $0$), $h_{d,u}$ is the inter-MT interference channel from the $u$-th UL MT to the $d$-th DL MT, and $h_{0,0}$ represents the residual SI channel seen at the receive antenna of the BS $0$ from its own transmit antenna due to the DL transmission. Assuming that the $u$-th UL MT and the BS are simultaneously transmitting, the baseband received signals at the BS $0$ and at the $d$-th DL MT can be mathematically expressed as
\begin{align}
y_{0} & \triangleq \sqrt{p_{\setU}} h_{0,u} s_{u} + \sqrt{p_{0}} h_{0,0} s_{0} + n_{0}, \label{Eq:Received_BS} \\
y_{d} & \triangleq \sqrt{p_{0}} h_{d,0} s_{0} + \sqrt{p_{\setU}} h_{d,u} s_{u} + n_{d}, \label{Eq:Received_DL}
\end{align}
respectively, where $s_{\ell}$ denotes the unit power data symbol transmitted by node $\ell$ and $n_{k}$ is the additive complex white Gaussian noise (AWGN) with variance $\sigma_{k}^{2}$ at the receiving node $k$. The second summands in \eqref{Eq:Received_BS} and \eqref{Eq:Received_DL} represent the interference terms due to the BS FD operation mode (namely, the SI and the inter-MT interference); in the HD mode, both terms vanish. 

\section{User Scheduling and Power Allocation} \label{sec:User_Scheduling_Schemes}
In this section, we first consider the case where fixed transmit powers are utilized for the concurrent UL and DL transmissions, and we propose three low complexity user scheduling algorithms. Then, we derive the optimal power allocation (OPA) between the UL and DL communications for maximizing the sum rate, which we exploit and incorporate in the user scheduling procedure to enhance its performance.

\subsection{User Scheduling Algorithms} \label{sec:Algos}
Assuming fixed transmit powers $p_{0}$ and $p_{\setU}$, and that all involved channel gains can be accurately estimated during an appropriately designed control plane, we present the following three low complexity user scheduling algorithms for the FD system under consideration.
\begin{itemize}
\item[\ref{alg:alg1}.] \textit{Received Signal Strength at both UL and DL (RSS-UL RSS-DL)}: The UL MT $u^{\star}$ and the DL MT $d^{\star}$ are selected as the ones with the highest channel gains in $\setU$ and in $\setD$, respectively (see Algorithm~\ref{alg:alg1}).
\end{itemize}

\begin{algorithm}[!t] \caption{\hspace{-1mm}\textbf{:} RSS-UL RSS-DL User Scheduling} \label{alg:alg1}
\smallskip
\begin{algorithmic}


\STATE \hspace{-4mm} \texttt{Step 1 \hspace{0.1mm}:} Select the UL MT $u^{\star}$ such that:
\begin{align} \label{eq:A1} \tag{$\mathrm{A1.1}$}
u^{\star} = \argmax_{u \in \setU}|h_{0,u}|^2.
\end{align}

\STATE \hspace{-4mm} \texttt{Step 2 \hspace{0.1mm}:} Select the DL MT $d^{\star}$ such that:
\begin{align} \label{eq:A2} \tag{$\mathrm{A1.2}$}
d^{\star} = \argmax_{d \in \setD}|h_{d,0}|^2.
\end{align}
\end{algorithmic} \vspace{-1.5mm}
\end{algorithm} \vspace{-1mm}

\begin{itemize}
\item[\ref{alg:alg2}.] \textit{Received Signal Strength at the UL and Signal-to-Interference-plus-Noise Ratio at the DL (RSS-UL SINR-DL)}: The UL MT $u^{\star}$ is selected first as the one with the highest channel gain in $\setU$, and then the DL MT $d^{\star}$ is selected as the one experiencing the maximum SINR (see Algorithm~\ref{alg:alg2}).
\end{itemize}

\begin{algorithm}[!t] \caption{\hspace{-1mm}\textbf{:} RSS-UL SINR-DL User Scheduling} \label{alg:alg2}
\smallskip
\begin{algorithmic}


\STATE \hspace{-4mm} \texttt{Step 1 \hspace{0.1mm}:} Select the UL MT $u^{\star}$ as in \eqref{eq:A1}.

\STATE \hspace{-4mm} \texttt{Step 2 \hspace{0.1mm}:} Select the DL MT $d^{\star}$ such that:
\begin{align} \label{eq:B2} \tag{$\mathrm{A2.1}$}
d^{\star} = \argmax_{d \in \setD} \frac{p_{0}|h_{d,0}|^2}{p_{\setU}|h_{d,u^{\star}}|^2+\sigma_{d}^2}.
\end{align}
\end{algorithmic} \vspace{-1.5mm}
\end{algorithm} \vspace{-1mm}

\begin{itemize}
\item[\ref{alg:alg3}.] \textit{Received Signal Strength at the DL and Signal-to-Leakage-plus-Noise Ratio at the UL (RSS-DL SLNR-UL)}: The DL MT $d^{\star}$ is selected first as the one with the highest channel gain in $\setD$, and then the UL MT $u^{\star}$ is selected as the one yielding the maximum SLNR, i$.$e$.$, as the MT that simultaneously presents high UL channel gain and creates low inter-MT interference to the scheduled DL MT $d^{\star}$ (see Algorithm~\ref{alg:alg3}).
\end{itemize}

\begin{algorithm}[!t] \caption{\hspace{-1mm}\textbf{:} RSS-DL SLNR-UL User Scheduling} \label{alg:alg3}
\smallskip
\begin{algorithmic}


\STATE \hspace{-4mm} \texttt{Step 1 \hspace{0.1mm}:} Select the DL MT $d^{\star}$ as in \eqref{eq:A2}.

\STATE \hspace{-4mm} \texttt{Step 2 \hspace{0.1mm}:} Select the UL MT $u^{\star}$ such that:
\begin{align} \label{eq:C2} \tag{$\mathrm{A3.1}$}
u^{\star} = \argmax_{u \in \setU}\frac{p_{\setU}|h_{0,u}|^2}{p_{\setU}|h_{d^{\star},u}|^2+\sigma_0^2}.
\end{align}
\end{algorithmic} \vspace{-1.5mm}
\end{algorithm} 

\noindent \textbf{Remark:} In the Algorithm~\ref{alg:alg1}, the steps of selecting the UL MT $u^{\star}$ and the DL MT $d^{\star}$ do not affect each other and may, therefore, be performed independently. On the other hand, in the Algorithms~\ref{alg:alg2} and \ref{alg:alg3}, knowledge of the UL (respectively DL) MT channel gain is not required in the DL (respectively UL) scheduling step. In fact, once the UL MT $u^{\star}$ (respectively the DL MT $d^{\star}$) is selected, it suffices to know only $h_{d,u^{\star}}$ $\forall d\!\in\!\setD$ (respectively $h_{d^{\star},u}$ $\forall u\!\in\!\setU$) for the Algorithm~\ref{alg:alg2} (respectively for the Algorithm~\ref{alg:alg3}), instead of $h_{d,u}$ $\forall d\!\in\!\setD$ and $\forall u\!\in\!\setU$.

\subsection{Optimal Power Allocation (OPA)} \label{sec:PC}
The user scheduling algorithms presented above assume that the system always operates in FD mode. However, in certain system configurations, i$.$e$.$, for certain values of the intended channels and levels of the inter-MT interference, the system rate performance is maximized by operating in HD mode. For that, we hereinafter aim at determining the optimal transmit powers $p_{0}^{\star}$ and $p_{\setU}^{\star}$ for the BS $0$ and for the scheduled UL MT $u^{\star}$, which maximize the instantaneous sum rate. The power allocation between the UL and DL transmissions is assumed to take place after selecting $u^{\star}$ and $d^{\star}$ using any of the user scheduling algorithms described in Section~\ref{sec:Algos}.

Capitalizing on \eqref{Eq:Received_BS} and \eqref{Eq:Received_DL}, the SINRs at the BS $0$ and at the scheduled DL MT $d^{\star}$ are given by
\begin{align}
\label{Eq:gamma_0} \gamma_0 & \triangleq \frac{p_{\setU}|h_{0,u^{\star}}|^2}{p_{0}|h_{0,0}|^2+\sigma_0^2}, \\
\label{Eq:gamma_DL} \gamma_{d^\star} & \triangleq \frac{p_{0}|h_{d^{\star},0}|^2}{p_{\setU}|h_{d^{\star},u^{\star}}|^2+\sigma_{d^{\star}}^2},
\end{align}
respectively. Hence, the instantaneous rates of the UL communication (i$.$e$.$, of the link between the scheduled UL MT $u^{\star}$ and the BS $0$) and of the DL communication (i$.$e$.$, of the link between the BS $0$ and the scheduled DL MT $d^{\star}$), measured in bps/Hz, can be computed as $\mathsf{R}_{0} \triangleq \log_{2} \left(1+\gamma_0\right)$ and $\mathsf{R}_{d^{\star}} \triangleq \log_{2} \left(1+\gamma_{d^{\star}}\right)$, respectively. The instantaneous rate $\mathsf{R}$ of the considered FD system is defined as the cumulative rate of the DL and UL communications, i$.$e$.$, $\mathsf{R} \triangleq \mathsf{R}_{0} + \mathsf{R}_{d^{\star}}$. Using the latter definitions, we consider the OPA strategy that solves the following optimization problem:
\begin{align}\label{Eq:Maximize_FD_Rate}
\left(p_{0}^{\star},p_{\setU}^{\star}\right) \triangleq \argmax_{\substack{p_{0} \in [0, P_{0}]\\p_{\setU} \in [0, P_{\setU}]}} \mathsf{R}(p_{0},p_{\setU}),
\end{align}
where we have highlighted the dependence of $\mathsf{R}$ on the powers of the DL and UL transmissions. Solving \eqref{Eq:Maximize_FD_Rate} yields the OPA strategy summarized below in Theorem~\ref{th:OPA}; therein, we make use of the following function definitions for $x\geq0$:
\begin{align}
\zeta (x) & \triangleq \frac{|h_{0,u^{\star}}|^2 \sigma_{d^{\star}}^2}{x|h_{0,0}|^2+\sigma_0^2} - |h_{d^{\star},u^{\star}}|^2, \label{Eq:zeta} \\
\eta (x) & \triangleq \frac{|h_{d^{\star},0}|^2 \sigma_{0}^2}{x|h_{d^{\star},u^{\star}}|^2+\sigma_{d^{\star}}^2} - |h_{0,0}|^2. \label{Eq:eta}
\end{align}

\begin{theorem} \label{th:OPA} \rm{
Given $\zeta (p_{0})$ and $\eta (p_{\setU})$ obtained from \eqref{Eq:zeta} and \eqref{Eq:eta}, respectively, the OPA strategy is determined as:
\begin{itemize}
\item[\textit{i})] If $\zeta (P_{0}) \geq 0$ and $\eta (P_{\setU}) \geq 0$, then the optimal power allocation is given by $\left(p_{0}^{\star},p_{\setU}^{\star}\right)=\left(P_{0},P_{\setU}\right)$, i$.$e$.$, FD mode with maximum transmit powers is optimal;
\item[\textit{ii})] If $\zeta (P_{0}) < 0$ and/or $\eta (P_{\setU}) < 0$, then \eqref{Eq:Maximize_FD_Rate} is solved by either $\left(p_{0}^{\star},p_{\setU}^{\star}\right)=\left(P_{0},P_{\setU}\right)$ (i$.$e$.$, FD mode with maximum transmit powers), or $\left(p_{0}^{\star},p_{\setU}^{\star}\right)=\left(0,P_{\setU}\right)$ (i$.$e$.$, HD mode in the UL direction with maximum transmit power), or $\left(p_{0}^{\star},p_{\setU}^{\star}\right)=\left(P_{0},0\right)$ (i$.$e$.$, HD mode in the DL direction with maximum transmit power).
\end{itemize}}
\end{theorem}

\begin{IEEEproof}
See Appendix~\ref{sec:A_OPA}.
\end{IEEEproof}

\vspace{1mm}

The OPA in the considered FD system has a remarkably simple nature, i$.$e$.$, the power allocation maximizing the system rate performance is binary. This simple binary OPA is exploited in the scheduling procedure and is incorporated as an additional step in order to maximize the system performance by switching between the optimal operation modes, i$.$e$.$, HD or FD mode. For the cases where the OPA strategy corresponds to HD mode, either in the UL or DL direction, one can repeat the scheduling of the UL or DL MT so as to maximize the HD rate: this corresponds to selecting the UL or DL MT as in \eqref{eq:A1} or \eqref{eq:A2}, respectively, of Algorithm~\ref{alg:alg1}. The OPA strategy for user scheduling can be summarized as follows:
\begin{itemize}
\item[B1.] \textit{User Scheduling with OPA}: After selecting the UL MT $u^{\star}$ and the DL MT $d^{\star}$ with any of the user scheduling algorithms presented in Section~\ref{sec:Algos} using the maximum allowable transmit powers, the FD/HD mode of operation that maximizes the system rate performance is determined. If the OPA yields the HD mode in the UL (respectively in the DL), then the UL MT $u^{\star}$ (respectively the DL MT $d^{\star}$) is recomputed so as to maximize the UL (respectively the DL) TDD communication rate (see Enhancement of \ref{alg:alg1}--\ref{alg:alg3} on the top of this column).
\end{itemize}

\begin{algorithm}[!t] \caption*{\textbf{Enhancement of \ref{alg:alg1}--\ref{alg:alg3}:} User Scheduling with OPA} \label{alg:alg4}
\smallskip
\begin{algorithmic}
%
%

\STATE \hspace{-4mm} \texttt{Step 1 \hspace{0.1mm}:} \hspace{0.4mm}Run Algorithm~\ref{alg:alg1}, or \ref{alg:alg2}, or \ref{alg:alg3} to obtain 

\hspace{1.44cm} $d^{\star}$ and $u^{\star}$.

\STATE \hspace{-4mm} \texttt{Step 2 \hspace{0.1mm}:} \texttt{If} $\eta (P_0) \geq 0$ and $\zeta (P_{\setU}) \geq 0$, then
\begin{align*} 
\left(p_{0}^{\star},p_{\setU}^{\star}\right)=\left(P_{0},P_{\setU}\right).
\end{align*}

\hspace{1.47cm} \texttt{Else}
\begin{align*} 
\hspace{1.47cm}\left(p_{0}^{\star},p_{\setU}^{\star}\right) = \argmax_{\substack{p_{0} \in \{0, P_{0}\} \\ p_{\setU} \in \{0, P_{\setU}\}}} \mathsf{R}(p_{0},p_{\setU}).
\end{align*}

\hspace{2cm} \texttt{If} $p_{0}^{\star} = 0$, select $u^{\star}$ as in \eqref{eq:A1}. \vspace{2mm}

\hspace{2cm} \texttt{Else} \texttt{If} $p_{\setU}^{\star} = P_{\setU}$, select $d^{\star}$ as in \eqref{eq:A2}. \vspace{2mm}

\hspace{2cm} \texttt{End}
											
\hspace{1.4cm} \texttt{End}
\end{algorithmic}
\end{algorithm} \vspace{-1mm}

\section{Sum-Rate Analysis\\ over Rayleigh Fading Channels} \label{sec:Analysis}
The average sum rate of the considered FD system with any of the user scheduling Algorithms~\ref{alg:alg1}--\ref{alg:alg3} presented in the previous section is defined as 
\begin{align} \label{Eq:R_Algorithms}
\overline{\mathsf{R}} &\triangleq \overline{\mathsf{R}}_{0} + \overline{\mathsf{R}}_{d^{\star}} \\
\nonumber & = \int_{0}^{\infty} \log_2(1+x)\big(f_{\gamma_0}(x)+f_{\gamma_{d^\star}}(x)\big) \diff x \\
\nonumber & = \frac{1}{\log2}\int_{0}^{\infty} \frac{2-F_{\gamma_0}(x)-F_{\gamma_{d^\star}}(x)}{x+1} \diff x
\end{align}
where $\overline{\mathsf{R}}_{0}$ and $\overline{\mathsf{R}}_{d^{\star}}$ denote the average rates of the UL and DL communications, respectively, and notations $f_\gamma(\cdot)$ and $F_\gamma(\cdot)$ represent the probability density function (PDF) and the cumulative distribution function (CDF), respectively, of a random variable (RV) $\gamma$. In deriving the last step in \eqref{Eq:R_Algorithms} we have used the Stieltjes transform of the complementary CDFs.

In the following, we provide closed-form expressions of \eqref{Eq:R_Algorithms} for the Algorithms~\ref{alg:alg1} and~\ref{alg:alg2} under the following assumptions: \textit{i}) all channels apart from $h_{0,0}$ are subject to Rayleigh fading with unit mean squared amplitude value; \textit{ii}) the SI power $|h_{0,0}|^2$ is a constant real positive number; and \textit{iii)} all the DL MTs have the same AWGN variance, i$.$e$.$, $\sigma_{d}^{2}=\sigma_{\setD}^{2}$ $\forall d\!\in\!\setD$. Due to space limitations, we omit the rate analysis for Algorithms~\ref{alg:alg3} and the user scheduling with OPA, which will be considered in a longer version of this paper. Note that the following results can be straightforwardly extended to the case where $h_{0,0}$ is subject to Ricean fading \cite{Dua12}. Before proceeding with the average rate expressions, we introduce the function $\xi_n(x,y)$, with $x,y\geq0$ and integer $n>0$, defined as 
\begin{equation}\label{Eq:Psi_function}
\xi_n(x,y) \triangleq \frac{(-x)^{n-1}}{\Gamma(n)} \bigg( \sum_{k=1}^{n-1}\Gamma(k)(-x)^{-k} y^{-k} - e^{xy}{\rm Ei}(-xy) \bigg),
\end{equation}
where ${\rm Ei}(\cdot)$ denotes the exponential integral function \cite[Eq$.$ (8.211/1)]{Gra00}; observe that, for $n=y=1$, we simply have $\xi_1(x,1) = -e^{x} {\rm Ei}(-x)$. Furthermore, we introduce the parameter $w_{n} \triangleq (1-p_0/p_\setU)^{-n}$.

\begin{theorem} \label{th:alg1} \rm{
The average sum rate of the considered FD system obtained with the Algorithm~\ref{alg:alg1} is given by 
\begin{align}
\label{Eq:R_alg1} \overline{\mathsf{R}}_{\rm A1} \triangleq \ & \overline{\mathsf{R}}_{0} + \sum_{k=1}^{K_\setD}\binom{K_\setD}{k}\frac{\left(-1\right)^{k+1}}{\log2}\left(\frac{p_0}{p_0-kp_\setU}\right) \\
\nonumber & \times \left(\xi_1\left(\frac{k\sigma_{\setD}^2}{p_0},1\right)-\xi_1\left(\frac{k\sigma_{\setD}^2}{p_0},\frac{p_0}{kp_\setU}\right)\right)
\end{align}
where $\overline{\mathsf{R}}_{0}$ is obtained as
\begin{equation}\label{Eq:R_alg1_UL}
\overline{\mathsf{R}}_{0} \triangleq \sum_{k=1}^{K_\setU}\binom{K_\setU}{k}\frac{\left(-1\right)^{k+1}}{\log2}\xi_1\bigg(\frac{k\left(p_0|h_{0,0}|^2+\sigma_0^2\right)}{p_\setU},1\bigg).
\end{equation}}
\end{theorem}

\begin{IEEEproof}
See Appendix~\ref{sec:A_alg1}.
\end{IEEEproof} \vspace{1mm}

\begin{theorem} \label{th:alg2} \rm{
The average sum rate of the considered FD system obtained with the Algorithm~\ref{alg:alg2} is given by
\begin{align}
\label{Eq:R_alg2}
& \overline{\mathsf{R}}_{\rm A2} \triangleq \overline{\mathsf{R}}_{0} + \frac{1}{\log2}\sum_{k=1}^{K_\setD}\binom{K_\setD}{k}\left(-\frac{p_0}{p_\setU}\right)^{k}\bigg(\sum_{\ell=1}^k(-1)^{\ell}\bigg. \\
\nonumber & \times \bigg. w_{\ell}\xi_{k-\ell+1}\left(\frac{k\sigma_{\setD}^2}{p_0},\frac{p_0}{p_\setU}\right) \! + (-1)^{1-k}w_{k}\xi_1\left(\frac{k\sigma_{\setD}^2}{p_0},1\right) \! \bigg)
\end{align}
where $\overline{\mathsf{R}}_{0}$ is defined in \eqref{Eq:R_alg1_UL}.}
\end{theorem}

\begin{IEEEproof}
See Appendix~\ref{sec:A_alg2}.
\end{IEEEproof} 

\subsection{Sum-Rate Scaling and Multi-User Diversity}
We next investigate the scaling law of the sum rate achieved by the proposed user scheduling algorithms in order to show their multi-user diversity gain. The asymptotic growth of the sum rate as the number of MTs goes to infinity, with fixed transmit powers, can be formally shown using the extreme value theory \cite{Haan06}. Using standard tools (e$.$g$.$, Smirnov and Gnedenko theorems), we can easily show that the asymptotic limit distribution of the CDFs of both $\gamma_0$ and $\gamma_{d^{\star}}$ belongs to the Gumbel domain of attraction. Due to space limitation, we provide a less formal, yet insightful, analysis. We present results only for the Algorithm~\ref{alg:alg1}, as the sum-rate scaling of the Algorithms~\ref{alg:alg2} and \ref{alg:alg3} can be derived in a similar way.  

\begin{theorem} \label{th:asympt}\rm{
The sum rate of the considered FD system obtained with Algorithm~\ref{alg:alg1} when $K_\setD,K_\setU\rightarrow\infty$ is given~by
\begin{eqnarray}\label{Eq:Scaling}
\overline{\mathsf{R}}_{\rm A1} \approx \log(\log(K_\setD) \log(K_\setU)) + \log \left(\frac{p_\setU}{p_0|h_{0,0}|^2 + \sigma_0^2}\right) \!.
\end{eqnarray}
} 
\end{theorem}

\begin{IEEEproof}
See Appendix~\ref{sec:Asympt_alg1}.
\end{IEEEproof} \vspace{1mm}

\section{Simulation Results and Discussion} \label{sec:Num}
In this section, the performance of the proposed user scheduling algorithms is evaluated. We provide simulation results for the average rate under Rayleigh fading (results for ultra-dense networks are presented in \cite{Atz16EW}). We also evaluate numerically the analytical expressions for the average sum rate of the Algorithms~\ref{alg:alg1} and~\ref{alg:alg2} presented in Section~\ref{sec:Analysis}. 
\begin{figure}[t!] 
\centering
\includegraphics[width=0.45\textwidth]{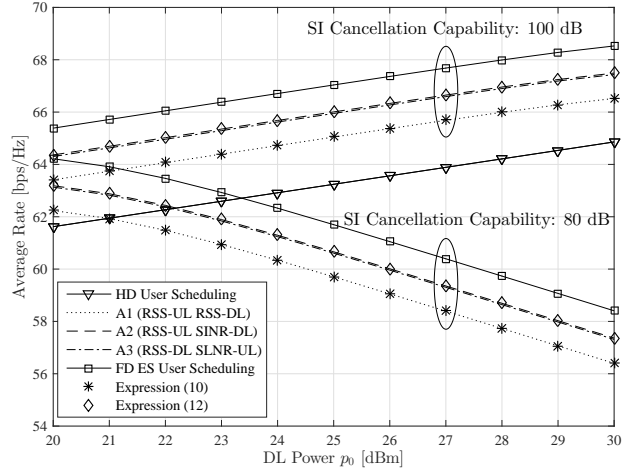}
\caption{Average sum rate of Algorithms~\ref{alg:alg1}--\ref{alg:alg3} vs$.$ the DL power $p_0$ for $p_\setU=0.95p_0$ dBm and $K_\setD=K_\setU=5$.}
\label{Fig:Figure_2} \vspace{-2mm}
\end{figure}

\begin{figure}[t!] 
\centering
\includegraphics[width=0.45\textwidth]{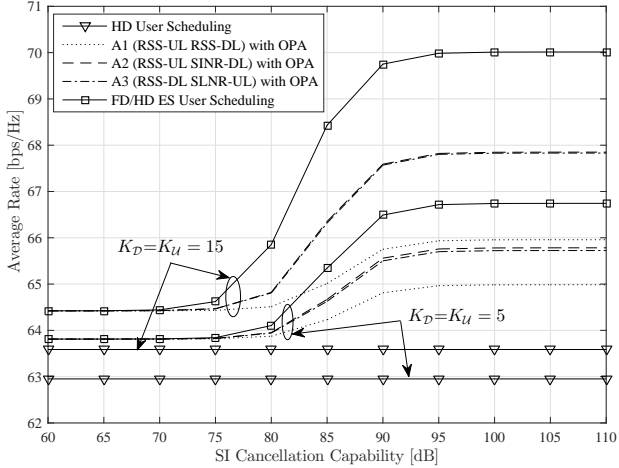} 
\caption{Average rate of Algorithms~\ref{alg:alg1}--\ref{alg:alg3} with OPA vs$.$ the SI cancellation capability for $p_0=24$ dBm and $p_\setU=23$~dBm.}
\label{Fig:Figure_3} \vspace{-2mm}
\end{figure}

In Figs.~\ref{Fig:Figure_2} and~\ref{Fig:Figure_3}, the average rate of the Algorithms~\ref{alg:alg1}--\ref{alg:alg3} and their OPA-enhanced versions is depicted for different values of $p_0$ and $p_\setU$, numbers of MTs in sets $\setD$ and $\setU$, and SI cancellation capabilities. For these results, we have used the typical values of $9$~dB noise figure for all DL MTs and $13$~dB noise figure for the small/pico cell BS, and computed $\sigma_\setD^2$ and $\sigma_0^2$ accordingly. Therein, we also plot the rate of user scheduling in the HD mode in TDD, and that of two versions of user scheduling with exhaustive search (ES): \textit{i}) the FD version that schedules concurrently one UL and one DL MT yielding the maximum FD sum rate; and \textit{ii}) the FD/HD version that schedules either the MT pair of the FD version or only the best DL MT or only the best UL MT, depending on which of the latter three modes provides the maximum rate performance. Firstly, it is shown in Fig.~\ref{Fig:Figure_2} that the numerically evaluated results for \eqref{Eq:R_alg1} and \eqref{Eq:R_alg2} match perfectly with their equivalent simulations, thus validating our analysis. In both figures, we observe that \ref{alg:alg2} and \ref{alg:alg3} exhibit similar sum-rate performance, which is expectedly always superior than that of \ref{alg:alg1}. This trend holds for both the original versions of the algorithms and their OPA-enhanced versions, and is mainly due to the fact that \ref{alg:alg1} ignores the inter-MT interference created after its first step. As shown in Fig.~\ref{Fig:Figure_2} for $K_\setD=K_\setU=5$, when the SI cancellation capability drops from $100$ to $80$~dB, increasing $p_0$ and $p_\setU$ decreases the sum rate of the FD mode of operation. In particular, for $p_0\geq23.5$~dBm and $p_\setU\geq22.3$~dBm, the performance of the optimal FD user scheduling becomes even lower than that of the HD mode. On the other hand, it is obvious from Fig.~\ref{Fig:Figure_3} that the OPA-enhanced versions of the Algorithms~\ref{alg:alg1}--\ref{alg:alg3} always outperform the rate of the HD user scheduling. In fact, their performance superiority over HD increases as the SI cancellation capability improves and/or the multi-user diversity increases. It was found from additional experiments not included here due to space limitation that, for SI cancellation capability equal to $80$ dB, the probability of the FD mode increases from $9\%$ to $26\%$ for $K_\setD=K_\setU=5$ to $K_\setD=K_\setU=15$. When the SI cancellation capability improves to $90$ dB, this increase is from $82\%$ for $K_\setD=K_\setU=5$ to $99\%$ for $K_\setD=K_\setU=15$. In Fig.~\ref{Fig:Figure_3}, it is depicted that the performance gap between the FD/HD ES user scheduling and each of the Algorithms~\ref{alg:alg1}--\ref{alg:alg3} with OPA increases as $K_\setD$ and $K_\setU$ increase, and/or the SI cancellation capability improves, reaching a maximum value when the sum rate of the optimal FD user scheduling always outperforms the rate of the optimal HD user scheduling. For example, for SI cancellation capability equal to $110$~dB and considering the OPA-enhanced versions of \ref{alg:alg2} and~\ref{alg:alg3}, this gap is $0.96$~bps/Hz for $K_\setD=K_\setU=5$ and $2.16$~bps/Hz for $K_\setD=K_\setU=15$.

\begin{figure}[t!] 
\centering
\includegraphics[width=0.45\textwidth]{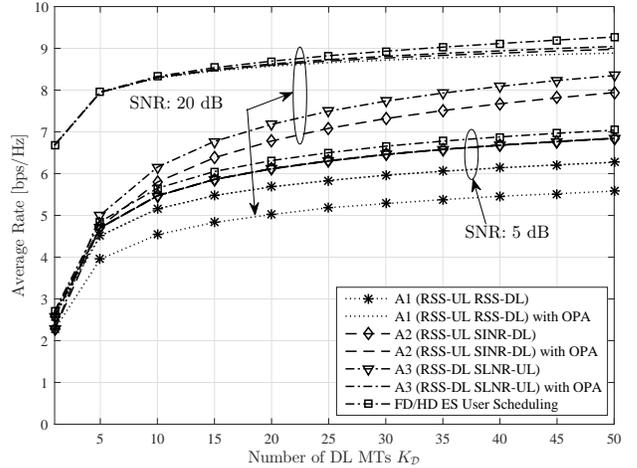}
\caption{Average rate of Algorithms~\ref{alg:alg1}--\ref{alg:alg3} and their OPA enhancements vs$.$ the number of DL MTs $K_\mathcal{D}$ for $K_\setU=K_\setD$ and SI cancellation capability of $20$ dB.}
\label{Fig:Figure_4} \vspace{-2mm}
\end{figure}

Finally, in Fig.~\ref{Fig:Figure_4}, we plot the sum rate of the proposed algorithms with and without the OPA enhancement versus the number of DL MTs $K_\setD$ ($K_\setD=K_\setU$). The performance of the OPA-enhanced algorithms is very close to the optimal ES for both considered values of the signal-to-noise ratio (SNR), while the performance gap of user scheduling without OPA decreases for increasing $K_\setD$. This implies that, with a large number of MTs to select from, the FD mode outperforms the HD one. Otherwise stated, multi-user diversity can compensate for the SI and the inter-MT interference due to the FD operation and boost the FD sum-rate performance.

\section{Conclusion} \label{sec:Conclusion}
In this paper, we investigated the problem of user scheduling and power control in FD cellular systems. We proposed three low complexity user scheduling algorithms to maximize the sum rate. We also derived the optimal power allocation for the two communication links and introduced efficient metrics for FD/HD mode switching as a means to further increase the system rate performance. For two of the proposed user scheduling algorithms for FD communication we presented closed-form expressions for their average sum-rate performance over Rayleigh fading. Our performance evaluation results unveiled that power control and low complexity user scheduling can boost the performance of FD small/pico cells. 

\appendices
\section{Proof of Theorem~\ref{th:OPA}} \label{sec:A_OPA}
\renewcommand{\theequation}{I.\arabic{equation}}
\setcounter{equation}{0}
The instantaneous sum rate of the considered FD system can be rewritten as $\mathsf{R} (p_{0},p_{\setU}) = \log_{2} \left( 1 + \varphi (p_{0},p_{\setU}) \right)$,
where we have defined the following real positive function for $x,y\geq0$:
\begin{align}
\hspace{-1mm} \varphi (x,y) \triangleq \ & \frac{x|h_{d^{\star},0}|^2}{y|h_{d^{\star},u^{\star}}|^2+\sigma_{d^{\star}}^2} + \frac{y|h_{0,u^{\star}}|^2}{x|h_{0,0}|^2+\sigma_0^2} \\
\nonumber & \hspace{-0.8mm} + \frac{xy|h_{d^{\star},0}|^2 |h_{0,u^{\star}}|^2}{\big(y|h_{d^{\star},u^{\star}}|^2+\sigma_{d^{\star}}^2 \big) \big(x|h_{0,0}|^2+\sigma_0^2 \big)}.
\end{align}
Recall the definitions of $\zeta (p_{\setU})$ and $\eta (p_{0})$ in \eqref{Eq:zeta} and in \eqref{Eq:eta}, respectively. From $\frac{\partial \varphi (p_{0},p_{\setU})}{\partial p_{\setU}}$ and $\frac{\partial^{2} \varphi (p_{0},p_{\setU})}{\partial p_{\setU}^{2}}$, it is straightforward to show that:
\begin{itemize}
\item[\textit{a})] If $\zeta (p_{\setU}) \geq 0$, then $\varphi (p_{0},p_{\setU})$ is strictly increasing and concave in $p_{0}$ and $\argmax_{p_{0} \in [0,P_{0}]} \mathsf{R} (p_{0},p_{\setU}) = P_{0}$;
\item[\textit{b})] If $\zeta (p_{\setU}) < 0$, then $\varphi (p_{0},p_{\setU})$ is strictly convex in $p_{0}$ and $\argmax_{p_{0} \in [0,P_{0}]} \mathsf{R} (p_{0},p_{\setU}) \in \{ 0, P_{0} \}$.
\end{itemize}
Likewise, from $\frac{\partial \varphi (p_{0},p_{\setU})}{\partial p_{\setU}}$ and $\frac{\partial^{2} \varphi (p_{0},p_{\setU})}{\partial p_{\setU}^{2}}$, it is not difficult to conclude that:
\begin{itemize}
\item[\textit{c})] If $\eta (p_{0}) \geq 0$, then $\varphi (p_{0},p_{\setU})$ is strictly increasing and concave in $p_{\setU}$ and $\argmax_{p_{\setU} \in [0,P_{\setU}]} \mathsf{R} (p_{0},p_{\setU}) = P_{\setU}$;
\item[\textit{d})] If $\eta (p_{0}) < 0$, then $\varphi (p_{0},p_{\setU})$ is strictly convex in $p_{\setU}$ and $\argmax_{p_{\setU} \in [0,P_{\setU}]} \mathsf{R} (p_{0},p_{\setU}) \in \{ 0, P_{\setU} \}$.
\end{itemize}
Putting all above together, we have shown that $p_{0}^{\star} \in \{ 0, P_{0} \}$ and $p_{\setU}^{\star} \in \{ 0, P_{\setU} \}$. Hence, one can check directly the conditions on $\eta (P_{0})$ and $\zeta (P_{\setU})$, as stated in Theorem~\ref{th:OPA}, to determine the OPA strategy. This completes the proof.

\section{Proof of Theorem~\ref{th:alg1}} \label{sec:A_alg1}
\renewcommand{\theequation}{II.\arabic{equation}}
\setcounter{equation}{0}
Starting from the DL communication, the CDF of the DL SINR $\gamma_{d^\star}$ in \eqref{Eq:gamma_DL} for the Algorithm~\ref{alg:alg1} can be obtained as
\begin{align}
\label{Eq:CDF_gamma_d_alg1} F_{\gamma_{d^\star}}(x) & = \hspace{0.8mm} \Pr\left[|h_{d^\star,0}|^2<\frac{p_\setU|h_{d^\star,u^\star}|^2+\sigma_{\setD}^2}{p_0}x\right] \\
\nonumber & \stackrel{(\rm{a})}{=} \int_{0}^{\infty} \Pr\left[\kappa<\frac{p_\setU y+\sigma_{\setD}^2}{p_0}x\bigg|y\right]f_{\lambda}(y) \diff y \\
\nonumber & \stackrel{(\rm{b})}{=} \int_{0}^{\infty} \left(1-\exp\left(-\frac{p_\setU y+\sigma_{\setD}^2}{p_0}x\right)\right)^{K_\setD}\!f_{\lambda}(y) \diff y,
\end{align}
where in $(\rm{a})$ we condition on $\lambda\triangleq|h_{d^\star,u^\star}|^2$ and we use the fact that $\lambda$ and $\kappa\triangleq\max_{d\in\setD}|h_{d,0}|^2$ are independent, and in $(\rm{b})$ we use the CDF of the maximum of $K_\setD$ independent and identically distributed (i$.$i$.$d$.$) exponential RVs. By applying the binomial expansion and using \cite[Eq$.$ (3.381/4)]{Gra00} to solve the resulting integrals, $F_{\gamma_{d^\star}}(x)$ can be derived in closed form as  
\begin{equation}\label{Eq:CDF_gamma_d_alg1_final}
F_{\gamma_{d^\star}}(x) = \sum_{k=0}^{K_\setD}\binom{K_\setD}{k} \! \left(-1\right)^k \! \left(k\frac{p_\setU}{p_0}x+1\right)^{-1} \! \!\exp\left( \! -k\frac{\sigma_{\setD}^2}{p_0}x\right).
\end{equation}

The CDF of the UL SINR $\gamma_0$ in \eqref{Eq:gamma_0} for the Algorithm~\ref{alg:alg1} (using the UL MT selection criterion in \eqref{eq:A1}) can be derived in a similar way to \eqref{Eq:CDF_gamma_d_alg1}, and results in
\begin{align}
\label{Eq:CDF_gamma_0_alg1} F_{\gamma_0}(x) & = \Pr\left[\max_{u\in\setU}|h_{0,u}|^2<\frac{p_0|h_{0,0}|^2+\sigma_0^2}{p_\setU}x\right] \\
\nonumber & = \sum_{k=0}^{K_\setU}\binom{K_\setU}{k}\left(-1\right)^k\exp\left(-\frac{p_0|h_{0,0}|^2+\sigma_0^2}{p_\setU}kx\right).
\end{align}
By substituting \eqref{Eq:CDF_gamma_d_alg1} and \eqref{Eq:CDF_gamma_0_alg1} into \eqref{Eq:R_Algorithms}, and then applying the integration theory of rational functions \cite[Sec$.$ 2.102]{Gra00} and using \cite[Eq$.$ (3.352/4)]{Gra00}, the average sum rate with the Algorithm~\ref{alg:alg1} is obtained as in \eqref{Eq:R_alg1} after some algebraic manipulations.

\section{Proof of Theorem~\ref{th:alg2}} \label{sec:A_alg2}
\renewcommand{\theequation}{III.\arabic{equation}}
\setcounter{equation}{0}
The selection criterion of the UL MT with the Algorithm~\ref{alg:alg2} is the same as with the Algorithm~\ref{alg:alg1} and, hence, we have the same average rate for the UL communication (cf$.$ expression \eqref{Eq:R_alg1_UL}). Focusing then on the DL communication, the CDF of DL SINR $\gamma_{d^{\star}}$ in \eqref{Eq:gamma_DL} for the Algorithm~\ref{alg:alg2} (using the DL MT selection criterion in \eqref{eq:B2}) can be obtained as
\begin{align}
\label{Eq:CDF_gamma_d_alg2} F_{\gamma_{d^\star}}(x) &= \Pr\left[\max_{d\in\setD}\frac{p_{0}|h_{d,0}|^2}{p_{\setU}|h_{d,u^\star}|^2+\sigma_{\setD}^2}<x\right] \\
\nonumber & \stackrel{(\rm{c})}{=} \left(\int_{0}^{\infty} \Pr\left[\tau<\frac{p_{\setU}y+\sigma_{\setD}^2}{p_0}x\bigg|y\right]f_{\nu}(y) \diff y\right)^{K_\setD} \\
\nonumber & \stackrel{(\rm{d})}{=} \sum_{k=0}^{K_\setD}\binom{K_\setD}{k}\left(-\frac{p_\setU}{p_0}x-1\right)^{-k}\exp\left(-\frac{\sigma_{\setD}^2}{p_0}kx\right),
\end{align}
where in $(\rm{c})$ we condition on $\nu\triangleq|h_{d,u^\star}|^2$ and use the fact that $\nu$ and $\tau\triangleq|h_{d,0}|^2$ are stochastically independent (for the proposed decoupled UL/DL user scheduling), and $(\rm{d})$ follows after substituting the marginal PDF of an exponential RV, then using \cite[Eq$.$ (3.381/4)]{Gra00}, and finally applying the binomial expansion. By substituting \eqref{Eq:CDF_gamma_d_alg2} and \eqref{Eq:CDF_gamma_0_alg1} into \eqref{Eq:R_Algorithms}, and then applying \cite[Sec$.$ 2.102]{Gra00} and using \cite[Eq$.$ (3.353/2)]{Gra00}, the average sum rate with the Algorithm~\ref{alg:alg2} is obtained as in \eqref{Eq:R_alg2} after some algebraic manipulations.

\section{Proof of Theorem~\ref{th:asympt}} \label{sec:Asympt_alg1}
\renewcommand{\theequation}{IV.\arabic{equation}}
\setcounter{equation}{0}
The scaling result in \eqref{Eq:Scaling} for the sum rate with the Algorithm~\ref{alg:alg1} is based on the following lemma for the Jensen's inequality for the logarithmic function.
\begin{lemma}\rm{
Let $\setI\triangleq\{1,2,\ldots,K\}$ and $Y_i$ $\forall i\!\in\!\setI$ be a sequence of positive i$.$i$.$d$.$ RVs. Also, let $Z_K\triangleq\max_{i \in \setI}Y_i$ with finite mean $\mu_K$ and finite variance $\sigma_K^2$, and $\Exp\left[\log(Z_K)\right] < \infty$. If $\displaystyle \lim_{K\rightarrow\infty}\!\!\sigma_K^2\mu_K^{-1}\!\to\!0$, then $\log\mu_K-\mathbb{E}[\log Z_K]\!\rightarrow\!0$ as $K\!\rightarrow\!\infty$.}
\end{lemma} 
Since $|h_{0,u}|^2$ $\forall u\!\in\!\setU$ are i$.$i$.$d$.$ exponential RVs, the mean and variance of the RV $Z_{K_\setU}\triangleq\max_{u \in \setU}|h_{0,u}|^2$ for $K_\setU \to \infty$ are given by $\mu_{K_\setU} = \setH_{K_\setU} \sim \log(K_\setU) + \gamma$ and $\sigma^2_{K_\setU} \sim \pi^2/6$, respectively, where $\mathcal{H}_{K_\setU}$ is the $K_\setU$-th harmonic number and $\gamma \cong 0.577$ is the Euler-Mascheroni constant \cite{Haan06}. Hence, we have that $\displaystyle \lim_{K_\setU\rightarrow\infty}\!\!\sigma_{K_\setU}^2\mu_{K_\setU}^{-1}\!\to\!0$; therefore, it holds for $\overline{\mathsf{R}}_0$ that 
\begin{eqnarray}
\overline{\mathsf{R}}_0 \approx \log(\log(K_\setU)) + \log\left(\frac{p_\setU}{p_0|h_{0,0}|^2+\sigma_0^2}\right).
\end{eqnarray}
Using similar tools from the extreme value theory, we can show that $\overline{\mathsf{R}}_{d^{\star}} \approx \log(\log(K_\setD))$, which concludes the proof.

\addcontentsline{toc}{chapter}{References}
\bibliographystyle{IEEEtran}
\bibliography{IEEEabrv,references}

\end{document}